\documentclass[prl,aps,groupaddress,showpacs,prd,twocolumn]{revtex4}
\usepackage{epsfig,amsmath,graphicx,amssymb}
\def\be{\begin{equation}}
\def\ee{\end{equation}}
\def\bea{\begin{eqnarray}}
\def\eea{\end{eqnarray}}
\def\bes{\begin{subequations}}
\def\ees{\end{subequations}}

\newcommand{\PT}{\mathcal{PT}}
\newcommand{\p}{\mathcal{P}}
\newcommand{\T}{\mathcal{T}}

\newcommand{\tOmega}{\tilde{\Omega}}

\begin{document}

\title{ $\PT$ symmetry with a system of three-level atoms}

\author{
Chao Hang$^1$, Guoxiang Huang$^1$, and Vladimir V. Konotop$^{2}$}
\affiliation{$^1$ State Key Laboratory of Precision Spectroscopy
and Department of Physics, East China Normal University, Shanghai
200062, China
\\
$^2$Centro de F\'isica Te\'orica e Computacional and
Departamento de F\'isica, Faculdade de Ci\^encias, Universidade de
Lisboa, 
Avenida Professor Gama Pinto 2, Lisboa 1649-003, Portugal }

\begin{abstract}
We show that a vapor of multilevel atoms driven by far-off
resonant laser beams, with possibility of interference of two
Raman resonances, is highly efficient for creating parity-time
($\PT$) symmetric profiles of the probe-field refractive index, whose
real part is symmetric and imaginary part is anti-symmetric in
space. The spatial modulation of the susceptibility is achieved by
proper combination of standing-wave strong control fields and of
Stark shifts induced by a far-off-resonance laser field.  As
particular examples we explore a mixture of isotopes of Rubidium
atoms and design a $\PT$-symmetric lattice and a parabolic
refractive index with a linear imaginary part.

\end{abstract}


\maketitle



While non-Hermitian operators obeying pure real spectra, like for
example the Bogoliubov-de Gennes~\cite{Bogoliubov} spectral
problem, linear stability problem for nonlinear waves, or simply a
parabolic potential with linear imaginary part~\cite{Kato}, are
known in physics for long time, it was only due to the work
~\cite{Bender}
that fundamental importance of
such operators became widely recognized. It was discovered
in~\cite{Bender} that there exists a wide class of complex
potentials of Schr\"odinger equation obeying pure real spectra,
and even most importantly, that this property is intrinsically
related to the parity ($\p$) and time ($\T$) symmetries of
physical systems. This discovery triggered the
discussion~\cite{QM} on the fundamentals of quantum mechanics
whose axioms are based on Hermitian operators for observables.
Further growth of interest in the theory of parity-time ($\PT$)
symmetric potentials was originated by suggestions of
implementation of $\PT$ symmetry in a waveguide  with gain and
absorption~\cite{Muga}, which was based on the
analogy between quantum mechanics and paraxial optics where the
refractive index plays the role of the potential in the
Schr\"odinger equation. In optics $\PT$-symmetric refractive index
profiles (i.e. obeying gain and losses of a given geometry) has
been experimentally realized using process of four-wave mixing in
Fe-doped LiNbO$_3$ substrate~\cite{nonrecip}. The possibility of
optical realization of $\PT$ symmetric potentials motivated
various suggestions of practical applications, like nonreciprocal
wave propagation~\cite{nonrecip,nonrecip1}, implementation of
coherent perfect absorber~\cite{Longhi}, giant wave
amplification~\cite{KSZ}, etc. Experimental realization of
$\PT$-symmetry using plasmonics~\cite{BDL} and temporal simulation
of lattices using optical couplers~\cite{nature} were also
reported.

In this Letter we demonstrate the possibility of
practical
implementation of spatially distributed $\PT$-symmetric
refractive index, i.e. the one having the property
$n(x)=n^*(-x)$,
in vapors of
multi-level atoms driven by control fields with properly
chosen Raman resonances and by a far-off-resonant Stark field.

First we recall some recent achievements in
creation of large susceptibilities in atomic vapors controlled by
external laser beams. While such systems are intrinsically
dissipative, it was suggested in~\cite{Scully} and shown experimentally
in~\cite{Zibrov}, that using the destructive interference in
imaginary part of susceptibility, it is possible to achieve
large real refractive indexes keeping the absorption small enough.
Recently, the idea of using two far-off-resonant control fields
for realizing  high susceptibility with
nearly zero absorption of a probe field
was developed theoretically~\cite{Yavuz} and confirmed
experimentally~\cite{PUGY}.  An alternative way of achieving similar effect
was proposed in~\cite{BARK} where two $\Lambda$-systems were explored
in order to excite two Raman resonances (see also
Ref.~\cite{Simmmons}).

Since the above mentioned schemes use the interference of two
Raman resonances, one of which resulting in gain and another in
absorption, the imaginary part of probe-field susceptibility
appears as a non-monotonic function of the frequency with both
positive (gain) and negative (absorbing) domains. Even more
remarkable property
is the
possibility to design distributions where real and imaginary parts
of the susceptibility appear respectively as even and odd
functions of the probe-field frequency~\cite{Yavuz,PUGY,BARK,Simmmons}.
Since
for a monochromatic beam the change $\omega\to -\omega$
is equivalent to the change $t\to-t$
the last property can be viewed
as a time inversion symmetry. Thus our goal can be formulated as
completing this time symmetry by the symmetry in the coordinate space.

\begin{figure}
\centering
\includegraphics[width=0.7\columnwidth]{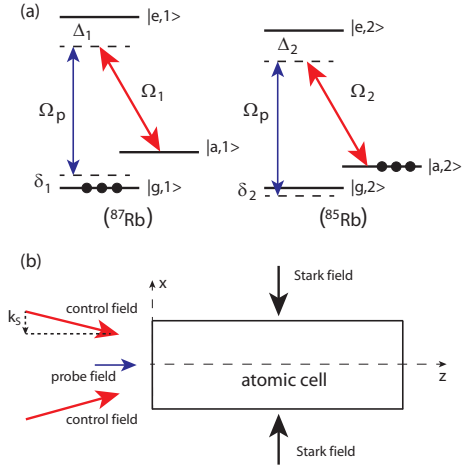}
\caption{(color online) (a) Two $\Lambda$-systems and Raman
transitions used for obtaining $\PT$-symmetric refractive index.
Initially
populated levels are indicated by the filled circles. (b)
Possible geometry for the suggested scheme.
All notations are defined in the text.
} \label{fig1}
\end{figure}

To achieve this goal we choose the scheme based on mixture of two
species of  $\Lambda$-atoms, similar to the
one explored in~\cite{BARK}. The involved atomic states will be
assigned as $|g,s\rangle$ (ground state), $|a,s\rangle$ (lower
state), and $|e,s\rangle$ (excited state), hereafter $s=1,2$ indicates
the specie of the atoms (Fig.~\ref{fig1}).
%
%
The species have atomic densities $N_1$ and $N_2$, respectively
($N=N_1+N_2$ being total atomic density). $\Omega_{p}$ is the half
Rabi frequency of the probe field coupling
$|g,s\rangle\leftrightarrow|e,s\rangle$ and $\Omega_{1,2}$ are
half Rabi frequencies of two control fields coupling
$|a,s\rangle\leftrightarrow|e,s\rangle$. All fields are far-off
resonant, which is guaranteed by condition
$\Delta_{s}\gg\Omega_{s}$, where
$\Delta_{s}=\omega_{e,s}-\omega_{a,s}-\omega_c$ is one-photon
detuning with $\hbar \omega_{l,s}$ ($l=g,a,e$) being the eigen
energy of the state $|l,s\rangle$, and $\omega_{p}$ ($\omega_{c}$)
being the center frequency of the probe (coupling) field.
Notice that one can explore different cases by changing
signs of the two-photon detuning, defined by
$\delta_s=\omega_{a,s}-\omega_{g,s}-(\omega_p-\omega_c)$. In
Fig.~\ref{fig1}(a)  we show an example of the proposed scheme
where $\delta_1>0$ and $\delta_2<0$, what corresponds to the $\PT$
symmetric lattice obtained below. In our
configuration,
the first scheme ($s=1$) exhibits two-photon absorption for the
probe field while the second one ($s=2$) provides two-photon gain.

Spatial modulation of
the probe-field susceptibility
can be achieved by using $x$-dependent
control-field [with Rabi frequency $\Omega_{s}(x)$].
Such filed, however affects both one- and
two-photon detunings. Therefore we explore the second possibility,
which is modulation of relative energy-level shifts along
$x$-direction resulting in the dependence
$\Delta_{s}=\Delta_{s}(x)$. This task can be achieved if a strong,
far-detuned laser field $E_{\rm S}(x)\cos(\omega_{\rm S} t)$,
where $E_{\rm S}$ and $\omega_{\rm S}$ are respectively amplitude
and frequency, is applied to the system. This field originates
Stark shifts of levels by $\Delta E_{{\rm S},s}(x) =
-\frac{1}{4} \alpha_{l,s} E_{\rm S}^2(x)$, where $\alpha_{l,s}$ is
the scalar polarizability of the level $|l,s\rangle$.  Below
$E_{\rm S}(x)$ is referred to as the Stark field. If within the
required accuracy one can consider
$\alpha_{g,s}\approx\alpha_{a,s}$, i.e. the difference of Stark
shifts between the ground-state sublevels is negligible (this is
the situation we consider below). Then $\delta_{s}$ is not
affected by the Stark field, while for the one-photon detunings we
have $\Delta_{s}(x)=\Delta_{s}- (\alpha_{e,s}-\alpha_{g,s})E_{\rm
S}^2(x)/(4\hbar)$. Thus the Stark field allows one to manipulate
the spatial distribution of $\Delta_{s}(x)$, on the one hand, and
on the other hand being far-off-resonant does not lead to power
broadening of lines due to low rate of
transitions~\cite{Letokhov}.

Since the Stark shift
usually appears in the second order of the perturbation theory, it
requires relatively strong electric fields. Taking into account
the characteristics of the available lasers, this  may impose
strong limitations, because the characteristic scale of the
spatial modulation of $\Delta_{s}(x)$ is of order of the
wavelength of the Stark field $\lambda_{\rm S}$. For the typical
order of the control field amplitudes  $E_c\sim 10^2$ V/cm
($\Omega_{1,2}\sim 2\pi\times 514$ MHz), the required amplitude of
the Stark field must be three orders of magnitude larger, i.e.
$E_{\rm S}\sim10^5$ V/cm (for more detail estimates see below).
Being focused into a spot of a diameter of $\approx 30$ $\mu$m
this requires laser powers of order of 100 W. Nowadays  such
powers can be achieved using say quantum cascade
lasers~\cite{QuantumLaser,review}
operating at micron wavelengths (in~\cite{QuantumLaser} it was $\lambda_{\rm
S}=4.45\,\,\mu$m, which will be used in our estimates).

For the described model the susceptibility of the probe field can
be computed from the density-matrix formalism within the
rotating-wave approximation. Its functional form is the same as
in~\cite{BARK} with the difference that now the two-photon
detuning and control field depend on the spatial coordinate:
\bea
\label{chi}
\frac{\chi_p(x)}{\chi_0}=
\frac{\delta_1-i\gamma_{ag}}{(\delta_1+\Delta_1(x)
-i\gamma_{eg})(\delta_1-i\gamma_{ag})-|\Omega_1(x)|^2}
\nonumber\\
-\eta \frac{|\Omega_2|^2
(\Delta_2(x)+i\gamma_{ag})^{-1}}{(\delta_2+\Delta_2(x)
-i\gamma_{eg})(\delta_2-i\gamma_{ag})-|\Omega_2(x)|^2}
. \eea
Here $\chi_0=N_1d_{eg,1}^2/(\varepsilon_0 \hbar)$, $\varepsilon_0$
is the vacuum permittivity, $\eta= N_2d_{eg,2}^2/N_1d_{eg,1}^2$
characterizing the ratio between the specie densities
is considered as a free parameter, the
$d_{..,s}$ stands for the dipole moment of the
transition between the ground an excited states of the $s$-the system.

Notice that generally speaking in a warm vapor
large Doppler broadened line widths
(typically $\Delta\omega/\omega\sim 10^{-6}$~\cite{Letokhov}) may degrade the effectiveness of resonant schemes.
While in our case such broadening would not be important for large one-photon detuning  it is of order of the two-photon detuning. It turns out, however, that the effect of  this line broadening can be significantly suppressed (by the factor $|\omega_c-\omega_p|/\omega_c$) if far-off-resonant co-propagating beams (see e.g.~\cite{broad}) with the frequencies close enough are used. This is the case we consider below.

Now our task
is to determine spatial
distributions of $\Omega_s(x)$ and  $\Delta_s(x)$
ensuring the condition
\be
\label{PT}
\nu(x)\equiv n(x)-n^*(x)=0,
\ee
where $n(x)=n_r(x)+in_i(x)\approx
[1+\frac{1}{2}\chi_p(x)]\sqrt{\varepsilon_0}$ is the refractive
index, for  $x$ being either arbitrary or belonging to some
interval (see
below).
$\nu(x)$ will be used to control the accuracy of the obtained $\PT$-symmetric refractive
indexes.

Because of the complexity of the relations among all parameters
involved in Eqs.~(\ref{chi}), (\ref{PT}),  it is not obvious {\em
a priori} that the problem has a solution.
 Therefore
 we
proceed with analysis of particular systems. To this end
we adopt the approach  as follows. First, we define a ``seed''
shape of the susceptibility, of the type we would like to obtain,
say $\chi^{sd}(x,\epsilon_j,\eta)$, which contains a number of
free parameters $\epsilon_j$ and $\eta$. Second, neglecting all
terms within some accuracy, say of 10\%, from (\ref{PT}) we
compute analytical solutions for
$\Omega_s^{sd}(x,\epsilon_j,\eta)$ and
$\Delta_s^{sd}(x,\epsilon_j,\eta)$. Third we substitute the
obtained ``seed'' Rabi frequencies and Stark shifts in
Eq.~(\ref{chi}). Due to crude approximations made the so obtained
susceptibility may still give significant errors in (\ref{PT}).
Our final step is to minimize the error using the control parameters
$\epsilon_{j}$ and $\eta$.

First, we apply the above algorithm to obtain a $\PT$-symmetric
lattice.
 Since the main limitation for the
lattice period is determined by the Stark field, we require the
period to be $\lambda_S$, i.e.  $\chi(x)=\chi(x+\lambda_S)$. The
experimental geometry we bear in mind is illustrated in
Fig.~\ref{fig1}(b). We further assume that each control field
consists two almost parallel plane waves having $x$-components of
the wavevector equal to $k_S=2\pi/\lambda_S$, i.e. to the Stark
field wavevector. Now the seed solution can be chosen as
$\chi^{sd}=\epsilon_0+\epsilon_1\cos\xi+i\epsilon_2\sin\xi $,
where $\xi=k_S x$ and $\epsilon_{0,1,2}$ are free parameters. As a
particular atomic vapor we use a mixture of isotopes $^{87}$Rb
($s=1$) and $^{85}$Rb ($s=2$), and assign $|g,s\rangle=|5S_{1/2},
F=1\rangle$, $|a,s\rangle=|5S_{1/2}, F=2\rangle$, and
$|e,s\rangle=|5P_{1/2}, F=1\rangle$ for each specie.

To perform numerical study we further particularize the problem by
assuming that $N_1=1.26\times10^{12}$ cm$^{-3}$ and $
N_2=1.14\times10^{12}$ cm$^{-3}$ (thus $N=2.4\times10^{12}$
cm$^{-3}$). The isotopes are loaded in a cell at
approximately $363\,$K and the coherence decay rates are
estimated as $\gamma_{eg}\approx\gamma_{ea}=2\pi\times334$ MHz and
$\gamma_{ag}=2\pi\times16$ kHz~\cite{BARK}. With sufficiently high
accuracy we can impose
$\alpha_{e,1}-\alpha_{g,1}\approx\alpha_{e,2}-\alpha_{g,2}=2\pi\hbar\times
0.1223$ Hz(cm/V)$^{2}$ and $ d_{ae,1}\approx d_{ae,2}=
2.5377\times10^{-27}$ C$\cdot$cm~{\cite{Steck}, what allows us to
consider $\Omega_1\approx\Omega_2=\Omega_c$. Other parameters are
chosen as $\omega_p\approx\omega_c=2\pi\times 3.77\times10^{14}$
s$^{-1}$ ($\lambda_p=795$ nm), $\eta=0.91$, $\chi_0=0.57$~s,
$\delta_{1}=1.80\gamma_{eg}$, and $\delta_{2}=-0.01\gamma_{eg}$.

As the second step, we solve
equation
$\chi_p(\Omega_s^{sd}, \Delta_s^{sd})=\chi^{sd}$ with respect the real $\Omega_s^{sd}$ and $\Delta_s^{sd}$. For $|\delta_2|\ll|\delta_1|$ it is
possible to leave only the leading terms in $\delta_2$ allowing one to
find the solutions explicitly. As the final step, we substitute $\Omega_s^{sd}$ and
$\Delta_s^{sd}$ back into Eq.~(\ref{chi})
and obtain $\nu(x)$ as small as possible by tuning the parameters
$\epsilon_{0,1,2}$. Since the obtained formulas have rather
cumbersome forms
they are not presented here. Instead, we write down the final expressions for the
control-field Rabi frequency and for the Stark-field amplitude by
keeping the first significant harmonics (i.e the terms $\gtrsim
10^{-4}$)
\bea
\begin{array}{l}
E_{\rm S}=E_0 [0.9698+0.0053\cos \xi -0.0007\sin \xi ],
\nonumber\\
\Omega_c=\gamma_{eg} [1.5384+0.0122\cos \xi +0.0232\sin \xi ].
\end{array}
\eea
Here $E_0=4\times 10^5$ V/cm [see
Fig.~\ref{fig2}(a)]. As we already mentioned, the final
shape of the susceptibility differs from the ansatz $\chi^{sd}$.
We represent it also in a form of a Fourier series by keeping the
first significant harmonics, i.e.
$\chi_p(x)=\chi_{0p}+\chi_{1p}(x)$, where $\chi_{0p}\approx
0.2257$ determines the average (in space) refractive index and the
periodic part is given by
\begin{equation}
\label{chi_data_1}
\chi_{1p}\approx 0.0075\cos\xi+i
10^{-4}[3.9418\sin\xi
+0.2298\sin (2\xi)].
\end{equation}
The real and imaginary parts of the refractive index is
respectively shown in Fig.~\ref{fig2}(b), (c). We notice that
there is a significant difference in the magnitude of the constant
real part of the susceptibility and its imaginary part. What is
important, however, is that the latter constitutes about several
percents of the variation of the real part. In Fig.~\ref{fig2}(d)
we show errors in obtaining the real and imaginary parts of
the susceptibility.
%
\begin{figure}
\centering
\includegraphics[width=\columnwidth]{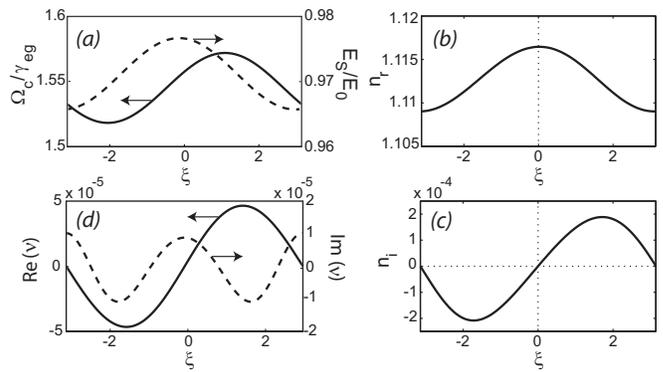}
\caption{ (a) The Rabi frequency of the control
(solid line) and
Stark
(dashed line) fields {\em vs.} $\xi$. (b) $n_r$ and (c) $n_i$ {\em vs.}
$\xi$. (d) Real (solid line) and imaginary (dashed line) parts of
the error function $\nu(\xi)$.} \label{fig2}
\end{figure}
%

Now turn to the propagation of the probe field along $z$-direction
in the described atomic vapor. Since in the $y$-direction the
medium is homogeneous, in the paraxial approximation the beam
propagation is governed by
\be\label{pro} 2ik_p\frac{\partial\Omega_p}{\partial z}
+\frac{\partial^2 \Omega_p}{\partial
x^2}+k_p^2\chi_{1p}(x)\Omega_p=0, \ee
where $k_p=\omega_p\sqrt{1+\chi_{0p}}/c$ is the probe-field
wavevector. [The beam of constant amplitude in $y$-direction is
chosen only for convenience: one can consider any waveguide
structure in $y$-direction what will result only in the
renormalization of the constants in (\ref{pro})].  By the  ansatz
$\Omega_p(x,z)=\tOmega(x)e^{ibz}$ where $b$ is the propagation
constant, we obtain
\be \label{eigen}
\frac{d^2\tOmega}{d\xi^2}+\frac{k_p^2}{k_S^2}\chi_{1p}(\xi)\tOmega= \beta
\tOmega, \quad\beta=\frac{2k_p}{k_S^2}b.
\ee
%
Taking into account the well known results on the periodic potentials~\cite{PT-lattice}
as well as rapid decay of the Fourier harmonics in (\ref{chi_data_1}),
one expects that with high accuracy the spectrum of (\ref{eigen}) is indeed pure real.
To check this, in Fig.~\ref{fig3}(a),(b) we show $\beta$ obtained numerically
for the susceptibility including all terms until the leading term violating the
condition (\ref{PT}) (the latter have amplitudes $\lesssim 10^{-5}$).

Let us now clarify whether the described procedure is
``structurally stable'', i.e. whether small deviations of system
parameters do not break the obtained $\PT$ symmetry.
To this end we tested the method with respect to
change of mutual concentration of the species. More specifically
we have changed the obtained $\eta$ by 10\% and repeated the
calculation described by the above algorithm (i.e. we
did not use $\eta$ as a matching parameter any more). We indeed
found that now the accuracy with which (\ref{PT}) is satisfied is
lower (see Fig.~\ref{fig3}(a),(b)\,)
%
\begin{figure}
\centering
\includegraphics[width=\columnwidth]{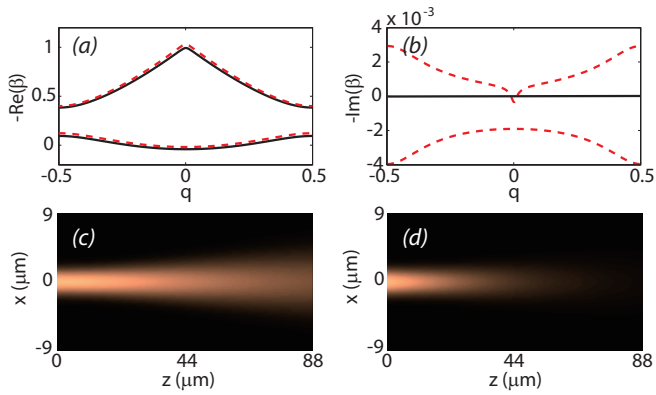}
\caption{(Color online) Real (a) and imaginary (b) parts of the dimensionless
propagation constant $\beta$. Solid (dashed) lines show the results for
$\PT$-symmetric refractive index and non $\PT$-symmetric refractive index
obtained by change of the mutual concentration of species, as explained
in the text. Evolution of $|\Omega_p(z)|^2$ in the vapors with $\PT$-symmetric
 (c) and non $\PT$-symmetric (d) refractive indexes.}
\label{fig3}
\end{figure}
%
but the imaginary parts are still very small (of order of
$10^{-3}$~\cite{com}). In Fig.~\ref{fig3} we also illustrate
propagation for the input Gaussian beam in the discussed
$\PT$-symmetric [panel (c)] and non-$\PT$-symmetric [panel (d)]
structures. The input probe beam
$\Omega_p(z=0)=e^{-0.1(k_Sx)^2}$ propagates
in the $\PT$-symmetric medium along much longer distance compared to the non $\PT$-symmetric case  where absorption is observed.

The described approach can be used to produce other
shapes of
the refractive index. As an example we now show how the method can be
modified to obtain a $\PT$ symmetric parabolic refractive index
~\cite{Kato}. The main idea is
based on the fact that $\PT$-symmetry can be satisfied only
locally in space. Then one can ``cut'' undesirable
non-$\PT$-symmetric distribution of the  refractive index by choosing
the location of the finite-size vapor cell with respect to the
domain where the external field produces local $\PT$-symmetry.
This is illustrated in Fig.~\ref{fig4} (a),(b) where we still consider the mixture of the
Rubidium isotopes, but taking now $^{85}$Rb as the first, $s=1$, and $^{87}$Rb as the second, $s=2$, systems. We choose $N_1=2.39\times10^{15}$ cm$^{-3}$ and $
N_2=8.98\times10^{12}$ cm$^{-3}$,
$\eta=0.375\times10^{-2}$,
$\delta_{1}=3.56\times10^{-4}\gamma_{eg}$, and
$\delta_{2}=3.93\times10^{-3}\gamma_{eg}$, without changing the other
parameters. The described procedure of the choice of the Stark and
control fields is performed to satisfy (\ref{PT}) only locally in
space. In particular, this is achieved, by taking $\Omega_c
=\gamma_{eg}(2+0.2\xi+0.009\xi^2)$ and
$E_{S}=E_{0}(0.3695-0.1375\xi)$, illustrated in Fig.~\ref{fig4}(c). The
spatial modulation of the susceptibility now has a complex
form [Fig.~\ref{fig4} a, b].
If however the atomic cell (limited say by Bragg mirrors) is
situated as shown in the panels, then the atomic vapor changes the
refractive index only in the domain occupied by the cell.
This domain was chosen to
ensure (\ref{PT}) with high accuracy where the real [solid line in panel (a)] and imaginary [solid line in panel (b)] are respectively parabolic and linear. Now the
susceptibility inside the atomic cell is described by
%
$ \chi_{p} = 0.2021-0.0007\xi^2 -i0.0006\xi$, which obvioulsy satisfies the condition
of  $\PT$ symmetry.
%
\begin{figure}
\centering
\includegraphics[width=\columnwidth]{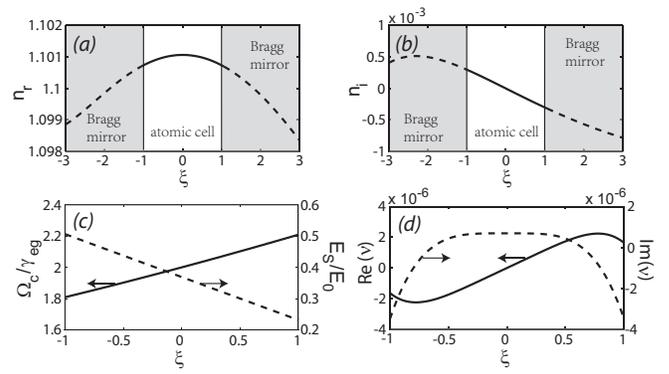}
\caption{Real (a) and imaginary (b) parts of
the
refractive index. Solid line is the refractive index inside the atomic
cell limited by Bragg mirrors.
Dashed
line shows the distribution if the gas would occupy the whole space. (c) Spatial distribution of Stark
(dashed
line) and control
(solid line)
fields required for inducing the local $\PT$-symmetric refractive index. (d)
Real (solid line) and imaginary (dashed line) parts of the error function inside the cell.
} \label{fig4}
\end{figure}
%

To conclude, we have suggested a possibility of creating
$\PT$-symmetric profiles of the refractive index in a mixture of
resonant atomic gases. An important property of the proposed
scheme is the possibility of changing parameters of the
structure {\it in situ}, say by changing wavelengths or amplitudes of
the control fields. While we used the
well studied isotopes of the Rubidium atoms the proposed scheme allows for further
generalization and improvement using other atomic isotopes,  four-level atoms,  mixtures
of more than two isotopes, monoatomic vapors with two control
fields, etc. Adding more control parameters opens
possibilities of experimental implementation of nonlinear
$\PT$-symmetric susceptibilities~\cite{nonlin}, as well as combined linear and
nonlinear ones~\cite{lin_nonlin} which recently attracted much attention.

 This work was supported by the Program of Introducing
Talents of Discipline to Universities under Grant
No. B12024, by NSF-China under Grant
Nos. 11174080 and 11105052, as well as
by the FCT grants
PEst-OE/FIS/UI0618/2011 and PTDC/FIS/112624/2009.


\begin{thebibliography}{99}
\bibitem{com}
Here the complex propagation constant is related to violation of (\ref{PT}) rather than to the $\PT$-symmetry breaking. With the accuracy $10^{-4}$ the optimized susceptibility has the form
$\chi_p=0.2298+0.0076\cos\xi+i[0.0026+10^{-4}(4.0083\sin\xi+1.8217\cos\xi)]$, i.e. its constant part results in the absorption of the light.
\end{thebibliography}
\end{document}